\documentclass[a4paper,twoside,12pt]{article}

\usepackage{graphicx}

\def \lsim{\mathrel{\vcenter
{\hbox{$<$}\nointerlineskip\hbox{$\sim$}}}}
\def \gsim{\mathrel{\vcenter
{\hbox{$>$}\nointerlineskip\hbox{$\sim$}}}}

\newcommand{\phys}{\rm {phys}}
\newcommand{\SUSY}{\rm {SUSY}}
\newcommand{\crit}{\rm {crit}}

\newcommand{\cosm}{\rm {cosm}}
\newcommand{\extra}{\rm {extra}}
\newcommand{\weak}{\rm {weak}}
\newcommand{\NP}{\rm {NP}}
\newcommand{\Strong}{\rm {Strong}}
\newcommand{\Planck}{\rm {Planck}}

\newcommand{\gev}{\rm{GeV}}
\newcommand{\tev}{\rm{TeV}}

\def \lsim{\mathrel{\vcenter
     {\hbox{$<$}\nointerlineskip\hbox{$\sim$}}}}
\def \gsim{\mathrel{\vcenter
     {\hbox{$>$}\nointerlineskip\hbox{$\sim$}}}}

\def\bea{\begin{eqnarray}}
\def\eea{\end{eqnarray}}
\def\be{\begin{equation}}
\def\ee{\end{equation}}
\def\ba{\begin{array}}
\def\ea{\end{array}}

\vspace{20pt}

\begin{document}

\begin{flushright}
CERN-PH-TH/2006-042\\
hep-ph/0607058
\end{flushright}

\hfill

\vspace{20pt}

\begin{center}
{\Large \textbf
{Physics Beyond the Standard Model}}
\end{center}

\vspace{40pt}

\begin{center}
\textsl{ \bf
 Riccardo Rattazzi} \vspace{12pt}

\textit{Department of Physics, CERN,
 CH-1211 Geneva 23, Switzerland}
\end{center}

\vspace{12pt}

\begin{center}
\textbf{Abstract }
\end{center}

\vspace{4pt} {\small \noindent 
I review recent theoretical work on electroweak symmetry breaking.}
\vskip 2.0truecm
{\it \noindent{International} Europhysics Conference on High Energy Physics\\
		 July 21st - 27th 2005\\
		 Lisboa, Portugal}
\vfill\eject 
\noindent

%
%






%

\section{Introduction}
I have been assigned this broad title but my talk will be mostly concerned with the origin of
the electroweak scale. I will attempt to give an overview of the theoretical `laborings' that came up  after the
end of the LEP era and in preparation to the commissioning of the LHC.
An appropriate subtitle for my talk could thus be `Electroweak Symmetry Breaking after LEP/SLC'.

There are two different sides from which to regard the legacy of LEP/SLC, and forming what is also known as the LEP paradox \cite{Barbieri:2000gf}.
From one side it is an impressive triumph of human endeavour: the Standard Model (SM)
is  a complete theory
of fundamental processes  successfully tested at the per-mille precision. That means  that small quantum corrections to the Born approximation  are  essential  in the comparison between theory and experiment.
However, when regarded from the other side, this great success becomes a huge conceptual bafflement, because 
the hierarchy problem, which inspired theoretical speculations for the last three decades, suggested that the SM should be overthrown right at the weak scale. That did not happen, so we must now understand why.
  I will discuss the paradoxical LEP/SLC legacy in the first part of my talk. 
In the second part I will attempt to give an overview on the new ideas that were stimulated partly by the LEP
paradox, on the phenomenological side, and partly by field theory developments (concerning for instance the use of extra-dimensions and branes) on the theoretical side. I think it will emerge that, while potentially realistic and certainly very ingenious, these attempts still leave something to be desired. In fact it may even be fair to say
that these models  concretely embody  the LEP paradox. Indeed, because of  the increasing
sense of frustration with the standard approaches,
a radically different approach to the hierarchy problem has  recently been advocated. That involves the
use of variants of the anthropic principle to explain the puzzling values of  apparently fundamental parameters, such as  the cosmological constant or the Higgs mass.
In the third part of my talk I will illustrate how anthropic considerations can explain the puzzling
need, after LEP2,  for tuning  on models with low energy supersymmetry.

\section{The legacy of LEP/SLC}
The large set of data collected in electron--positron collision at LEP/SLC displays
a remarkable $O(10^{-3})$ agreement with the SM for  a relatively light Higgs. More precisely, a global eletroweak fit  \cite{deJong:2005mk}  gives with 95\% CL the bound $m_h<219 \,\gev$. On the other hand, the SM suffers from the hierarchy problem: the Lagrangian Higgs
mass  parameter $m_H^2$, which is related to the physical mass by $m_h^2=-2 m_H^2$,
 is affected by uncalculable cut-off dependent quantum corrections. Whatever  more fundamental theory  replaces the SM above some scale $\Lambda_{\NP}$, barring
unwarranted cancellations, it is reasonable
to expect    the Higgs mass parameter to be at least of the same size as (or bigger than)  
the SM contribution computed with a cut-off scale $\Lambda_{\NP}$. (This way of estimating the size of the Higgs mass  is made reasonable by many explict examples  that solve the hierarchy problem, and also by analogy with well-known quantities in low energy physics, such as
the electromagnetic contribution to $m_{\pi^+}^2-m_{\pi^0}^2$.). 
The leading quantum correction  is then expected to come from the top sector and is estimated to be 
\be
\label{quadratic}
\delta m_H^2\sim -\frac{3\lambda_t^2}{8\pi^2}\Lambda_{\NP}^2\, .
\ee
In the absence of tuning, this contribution is compatible with the allowed range of $m_h^2$
only if the cut-off is rather low
\be
\label{notuning}
\Lambda_{\NP}< 600 \times (\frac{m_h}{200\, \gev})\, \gev\, .
\ee
If we allow a fine-tuning of order $\epsilon$ then the bound is relaxed by a factor $1/\sqrt{\epsilon}$. Now, the question is: if the energy  range of validity of the SM is as low as 
$500$--$1000\, \tev$, why did LEP/SLC not detect  any deviation from the SM predictions
in their rich set of data? Even though the center of mass energy of these experiments is significantly lower than $1\, \tev$, still their precision is high enough to make them sensitive
to new virtual effects associated to a much higher scale than their center of mass energy.
The effects from new physics at a scale $\Lambda_{\NP}$ can in general be parametrized by adding to the SM renormalizable Lagrangian the whole tower of  higher dimensional 
local operators, with coefficients suppressed by the suitable powers of $\Lambda_{\NP}$ \cite{Buchmuller:1985jz}:
\be
\label{effective}
{\cal L}_{eff}^{\NP} =\frac{1}{\Lambda_{NP}^2}\left \{c_1 (\bar e \gamma_\mu e)^2 +c_2 
W_{\mu\nu}^I B^{\mu\nu}H^\dagger\tau_I H+\dots\right\}\,.
\ee
At leading order it is also sufficient to consider only the operators of lowest dimension,
$d=6$. The constraints on the whole set of $d=6$ operators have been studied in ref.~\cite{Barbieri:1999tm}.  The lower bound on $\Lambda_{\NP}$ for each individual operator ${\cal O}_i$, neglecting the effects of all the others and normalizing $|c_i| = 1$, ranges between $2$ and $10$ TeV.  Turning  several coefficients on at the same time does not  qualitatively change
the result, unless parameters are tuned.
The interpretation of these results is that if New Physics affects electroweak observables at tree level, for which case $c_i\sim O(1)$, the generic lower bound on the new threshold is a few TeV.
The tension between this lower bound and eq.~(\ref{notuning}) defines what is known as the LEP paradox. This is an apparently mild problem. But notice that the needed tuning $\epsilon$ grows quadratically with $\Lambda_{\NP}$, so that for $\Lambda_{\NP} = 6$ TeV we need
to tune to 1 part in a hundred in order to have $m_H=200$ GeV. In view of this problem,
things would look definitely better if New Physics affected low energy quantities only via loop
effects. In this case $c_i\sim \alpha/4\pi$ and $\Lambda_{NP}<600$ TeV would not lead
to any  tension with electroweak precision tests. It is at first reassuring that Supersymmetry
with $R$-parity, arguably the leading candidate  New Physics scenario, precisely enjoys
this property, with the mass scale of supersymmetric particles identified with $\Lambda_{NP}$. However the attraction of Supersymmetry largely lies in its giving a very
plausible picture for physics way above the weak scale and up to the Planck scale where,
in addition to electroweak symmetry breaking (EWSB), also Gauge Unification, neutrino masses
and Dark Matter fit very well. In this extrapolation, however, the leading quantum
contribution to the Higgs mass parameter is not eq.~(\ref{quadratic})
but the larger one associated to renormalization group (RG) logarithms.  
In the Minimal Supersymmetric Standard Model (MSSM)
the relation between the various mass parameters is then roughly 
\bea
m_Z^2\,\sim -2 m_H^2&=& -2\mu^2 +\frac{3}{2\pi^2}\lambda_t^2 m_{\tilde t}^2 \ln \frac{M_{Planck}}{m_{\tilde t}}
+\dots \\
&\sim& -2\mu^2+O(1)\, m_{\tilde t}^2 +\dots\, ,
\label{zstop}
\eea
where we have not displayed the normally less relevant contributions.
From the above we deduce that the natural expectation is to have the
stop, the charginos and everything else at or below the vector boson scale
\be
m_Z\,\sim\,m_{\tilde t},\sim \mu\sim \dots\, 
\ee
The  above  relation raised great hope of new discoveries at  LEP/SLC.
This did not happen, and so supersymmetry can no longer be viewed as
completely natural. In fact, at least in the MSSM, 
the situation is made even worse by the indirect, and stronger, 
bound placed on the stop mass by the lower bound on the lightest Higgs mass.
As  is well know,  in the MSSM the physical mass of the  lightest Higgs
has an upper bound, which in 1-loop accuracy reads  roughly 
\be
\label{mh}
m_h^2 \leq m_Z^2 +m_t^2 \,\frac{3\lambda_t^2}{2\pi^2} \ln{m_{\tilde t}/m_t}\, .
\ee
The second term on the right-hand side corresponds to the leading top/stop radiative correction
to the  Higgs quartic coupling.
It is then only thanks to this correction that $m_h$ can exceed its direct experimental (95\% CL) lower bound of $114.4$ GeV. However, this generically requires 
$m_{\tilde t}\gsim 500$--$1000$ GeV, which when compared to eq.~(\ref{zstop})  implies
that a cancellation with $1$ to $5\%$ accuracy is needed. 
 Although the description we give here is somewhat schematic, the problem
is `robust', in the sense that it does not depend in any significant way on the full structure of the soft terms. In particular things are not dramatically improved by considering the extra positive contribution to the right-hand side of eq.~(\ref{mh}) that arises for large
$\tilde t_L$--$\tilde t_R$ mixing. This is because the sizeable $A$-terms that are needed for that to happen
require some tuning too. Another  often heard criticism to the above simple argument 
concerns the fact that the bound on $m_h$ in the MSSM is, strictly speaking, lower than $114.4$ GeV.
This is because the coupling of $h$ to the Z-boson is a factor $\sin ({\beta-\alpha})$ smaller than the one
 in the SM. In some regions of the supersymmetric parameter space this suppression can become significative, making the bound on $m_h$ weaker and thus giving the impression that the need for tuning is relaxed.
   However as a direct analysis shows \cite{romanino,agr},
the parameter space region where this happens corresponds to an even bigger tuning than the 
$1$--$5\%$ estimated above.
 This is because one needs  $\tan\beta \gg 1$ (which always entails some tuning), the  mass
 of the second CP-even Higgs $m_H$ tuned  somewhat close to $m_h$ and a still  sizeable
 stop contribution to the Higgs quartic coupling.
 
 While the problem is `robust' within the MSSM, it can be somewhat relaxed just by adding
 a single superfield $N$ to the model, thus upgrading the theory to the so-called NMSSM. 
 In the presence of $N$ there is an additional positive contribution to the right-hand side of eq.~(\ref{mh}), due to the superpotential trilinear coupling $NH_1H_2$. This allows a relaxation of the lower bound on the stop mass. A detailed analysis, described in ref.~\cite{Bastero-Gil:2000bw}, shows that the amount of fine tuning can in general be relaxed to about $10\%$. This  is encouraging, although my impression is that in several attractive scenarios for supersymmetry breaking, such as gauge or anomaly mediation,
 the soft terms have such a structure as to make the desired electroweak vacuum
 with $\langle H_2\rangle,\,\langle H_1\rangle,\,\langle N\rangle\not = 0$ rather hard to obtain,
 that is to say very tuned. Some extra model building effort in the context of the NMSSM is perhaps desirable.
 
 In the end, should we really  worry about  tuning at the few per cent level? Perhaps not, but
 we should  keep in mind that once we are willing to accept some tuning, the motivation for New Physics at the LHC becomes weaker. Notice indeed that, already with a tuning at the per mille level, the sparticles are out of reach at the LHC.

 \subsection{Technical parenthesis: LEP1 \& LEP2 bounds on New Electroweak Physics}
 I now want  to illustrate the impact of electron--positron data by focusing on the simplest scenario for New Physics in the electroweak sector, the so-called universal models. These are the models where deviations from the SM appear, at leading order, only through vector boson vacuum polarizations
 \cite{tech}
 \bea
 \label{oblique}
 {\mathcal L}_{NP} \,& = &\, W_+^\mu \Pi_{+-}(q^2)W_{+\mu}\,+\,W_3^\mu \Pi_{33}(q^2)W_{3\mu}\\
&+&\,W_3^\mu \Pi_{3B}(q^2)B_{\mu}\,+\,B^\mu \Pi_{BB}(q^2)B_{\mu}
\eea
Most Technicolor, Little Higgs and Higgsless models practically belong to this class \cite{Barbieri:2004qk}, showing that it is not an obviously idle exercise to focus on universality.  I say `practically',
since the
more realistic versions of these models almost always display extra effects involving the third-family fermions,
and associated to the large value of the top quark Yukawa coupling. However, since the majority of the observables (and arguably those that are under better experimental control) only 
involve the fermions of the first two families, the bounds on universal models  indeed have a more general relevance.

The electroweak constraints on universal models were widely discussed in the 90's.
However, as I will now show, and as it was  recently discussed in ref. \cite{Barbieri:2004qk}, some important aspects  were always either missed or not emphasized. Consistent with the absence of new particles at LEP2, let me start by assuming
that the scale of new physics $\Lambda_{\NP}$ is somewhat above the energy of LEP2. It  then makes  sense to expand the vacuum polarizations $\Pi(q^2)$ as a power series in $q^2$ and retain
only the leading terms. In order to decide which terms are leading, it is useful to classify
the vacuum polarizations in  eq.~(\ref{oblique}) according to their transformation properties
under custodial symmetry and under the electroweak group $SU(2)_L$ (the two relevant symmetries of the problem).  Within any given symmetry class is then natural to retain only the term of lowest order in the Taylor  expansion in $q^2$. This is because, barring accidental cancellations that make the lowest-order term in a given class anomalously small, the higher-order terms in the same class will give effects at around the $Z^0$ pole that are smaller by at least a factor $M_Z^2/\Lambda_{\NP}^2\ll 1$. According to this criterion, and after  reabsorbing the trivial redefinition of the electroweak input parameters $(G_F,\alpha_{EM},m_Z)$, we are left with 4 leading form factors
$$
\begin{array}{rclrlcc}
\multicolumn{3}{c}{\hbox{Adimensional form factors}}&
\multicolumn{2}{c}{\hbox{Operators}}& \hbox{Custodial} & \hbox{SU(2)$_L$}\\ \hline
{\widehat{S}} &=&g^2 \Pi'_{3 B}(0) & {\cal O}_{WB}~=&(H^\dagger \tau^a H) W^a_{\mu\nu} B_{\mu\nu} /gg'\!\!\!&+&-\\[1mm]
{\widehat{T} }&=& \frac{g^2}{M_W^2}\left (\Pi_{3 3}(0)-\Pi_{+-}(0)\right )\!\!\!
& {\cal O}_H~=&|H^\dagger D_\mu H|^2&-&-\\[1mm]
Y &=&\frac{g^{\prime 2}M_W^{2}}{2}\Pi''_{BB}(0) &{\cal O}_{BB}~=&(\partial_\rho B_{\mu\nu})^2/2g^{\prime 2}&+&+\\[1mm]
{W} &=&\frac{g^{2} M_W^{2}}{2} \Pi''_{3 3}(0) & {\cal O}_{WW} ~=&(D_\rho W^a_{\mu\nu})^2/2g^2&+&+\\[1mm]
\end{array}
$$
where we have indicated respectively with $+$ or $-$ the symmetries they respect or break.
We have also indicated the lowest dimension effective operator involving the Higgs
and vector fields associated to each form factor. As was already pointed out long ago by Grinstein and Wise \cite{Grinstein:1991cd}, the 4 leading form factors parametrize the $d=6$ effective Lagrangian for
the Higgs and gauge fields.
They are thus the leading terms in a double expansion in $\langle H\rangle^2/\Lambda_{\NP}^2$ and in $q^2/\Lambda_{\NP}^2$. It follows, however, from our discussion that they are the leading effects in 
full generality, as we did not assume we could expand in the Higgs field, and our parametrization also encompasses the generic strongly coupled  Higgsless scenario.
We stress that according to our criterion the quantity $
U \, = \, g^2\left (\Pi^\prime_{33}(0)-\Pi^\prime_{+-}(0)\right )$ is expected to be 
$\sim  \frac{m_W^2}{\Lambda_{NP}^2} \hat T\ll \hat T$ so that it  can always be safely  neglected.
The negligibility of $U$ is indeed a known property of technicolor models \cite{pt}. The quantities $Y$
and $W$ are also  small in the simplest technicolor models, but they can be important in models
where there is new structure in the pure gauge sector, as in models with vector boson compositeness or as in Little Higgs models.
On the other hand there exists, as expected,  no motivated scenario where  $S,T,U$ is the relevant set: it is either redundant or insufficient.

Notice that by the equations of motion the operators associated to $Y$ and $W$ are equivalent
to a given combination of $\widehat S,\widehat T$ plus vertex corrections and plus four-fermion contact interactions. 
Two classes of observables are then affected by $({\widehat S},{\widehat T}, Y, W)$:
\begin{enumerate}
\item $Z^0 $ pole. Corrections to $(\delta\rho\vert_{m_Z},\, m_W,\, \sin^2\theta_W\vert_{\mathrm {current}})$, expressed via the $\epsilon$'s of ref.\cite{alba} as
\bea
\epsilon_1\,&=&\,\epsilon_1^{SM}\,+\,\hat T\,-\,W\,-\,\tan^2\theta_W Y\\
\epsilon_2\,&=&\,\epsilon_2^{SM}\,-\,W\\
\epsilon_3\,&=&\,\epsilon_3^{SM}\,+\,\hat S\,-\,W\,-\, Y\, .
\eea
\item  Cross-sections and asymmetries in $e\bar e\,\to\, f\bar f$ at LEP2. These mostly constrain
$Y$ and $W$ since their effect grows faster with energy than that of ${\widehat S}$ and ${\widehat T}$ (they involve more derivatives).
\end{enumerate}
Notice that $Z^0$ pole tests correspond to the measurement of just 3 quantities, and are thus not sufficient
to constrain the general set! (As  is well known, the set $S,T,U$ would indeed be constrained
by $Z^0$ pole data: Is this the psychological reason why this inconsistent set was  so popular for so long?).
Fortunately LEP2 data allow us to fully and strongly constrain the set. It is interesting that
the somewhat lower precision of LEP2  (about $1\%$ versus about $0.1\%$ at LEP1) is compensated by the higher center of mass energy, which enhances the effect of $Y$ and $W$.  Other low energy
observables, such as atomic parity violation and Moeller scattering, also provide extra independent constraints,
but they are weaker than those provided by LEP2.
The bounds from the  global (basically LEP1/SLC + LEP2)  fit is shown in the table: all 4 quantities are bounded at the per-mille level.  The message should then be clear: LEP2 data are crucial to perform a consistent analysis of new electroweak physics.

\begin{table}
$$ \begin{array}{c|cccc}
\hbox{Type of fit} &10^3  \widehat S &  10^3\widehat T &10^3 Y & 10^3W \\ \hline
\hbox{One-by-one (light Higgs)} & \phantom{-}0.0\pm0.5& 0.1\pm0.6& 0.0\pm0.6& -0.3\pm0.6 \\
\hbox{One-by-one (heavy Higgs)} & \hbox{---} & 2.7\pm0.6&\hbox{---}&\hbox{---}\\ \hline
\hbox{All together (light Higgs)}& \phantom{-}0.0\pm 1.3 & 0.1 \pm 0.9 & 0.1 \pm 1.2 & -0.4 \pm 0.8 \\
\hbox{All together (heavy Higgs)} & -0.9\pm 1.3 & 2.0 \pm 1.0&0.0 \pm 1.2 & -0.2\pm 0.8 \\
\end{array}$$
\caption{\label{tab:fit}\em  Global fit (excluding NuTeV) of dominant form factors
including them one by one or all together, with a light ($m_h=115\, \gev$) and with a heavy ($m_h=800\, \gev$) Higgs.}
\end{table}

\section{`New' ideas on electroweak symmetry breaking}
Because of the `uncomputability' of  the Higgs potential, the SM, while a perfectly consistent theory,
does not give a satisfactory explanation of EWSB. Perhaps roughly:
the SM can parametrize EWSB but  cannot explain it. Sticking to theories with an elementary Higgs field,
progress necessarily involves  computational control of the Higgs mass parameter. That means that
$m_H^2$ should be protected from ultraviolet corrections. The only way we know of achieving this  is by introducing extra symmetries. There are various possibilities, by now well known.
Supersymmetry is surely the most widely explored one. By supersymmetry the Higgs boson $H$ 
is mass degenerate with a Higgs fermion $\Psi_H$ within the same Higgs supermultiplet. The Higgs
mass $m_H$ therefore inherits by supersymmetry the good UV property of the fermion mass:
the quadratic divergence is replaced by a mild logarithmic one, and the hierarchy problem is solved.
Another, perhaps less popular but interesting, possibility is to promote the Higgs to a gauge field.
We know indeed that  a gauge symmetry $\delta A_\mu=\partial_\mu\alpha$ forbids a mass term
$m^2A_\mu A^\mu$. In order for this to work the Higgs $H$ should be part of a vector multiplet, which
at first glance  conflicts with ordinary 4-dimensional Lorentz invariance.  However, the conflict is solved if there
exists (at least) one extra space dimension, in which case $H$ can be associated to the vector polarization along the new dimension: $H\sim A_5$. It is amusing that also supersymmetry can be viewed as an extra dimension,
though of fermionic type.
 Finally another, perhaps simpler, possibility is that the Higgs $H$ is in lowest
approximation the Goldstone boson of a spontaneously broken global symmetry. This means that $H$ basically transforms by  a constant shift $H\to H+c$ under the symmetry, which forbids any $H$ interaction that does not involve at least one derivative $\partial_\mu H$. In particular it forbids a Higgs mass term, but also,
which is less exciting, the standard Yukawa interactions and the Higgs self-coupling. In fact this is a more general problem: all the
symmetries I mentioned above must  be broken at some level in order to give rise to realistic models. Breaking the symmetry
while preserving its benefits, and also avoiding the LEP paradox, is the main challenge in model building. I will now illustrate some of these model building efforts. 

\subsection{The Little Higgs model}
The LEP paradox is overcome if we can construct a theory where $m_H$, with respect to $\Lambda_{NP}$,
is much smaller than eq.~(\ref{quadratic}) suggests. The Little Higgs (LH) idea is to achieve this construction by making 
the Higgs an approximate Goldstone boson (a pseudo-Goldstone in jargon) \cite{Kaplan:1983fs}. The inspiration for that
comes from low energy hadron physics, where the pions $\pi^+,\,\pi^0$ represent the Goldstone bosons
associated to the spontaneous breakdown of the chiral symmetry group $SU(2)_L\times SU(2)_R$ down
to the diagonal isospin group $SU(2)_I$. The quark masses $m_q$ and $\alpha_{EM}$ explicitly break 
chiral symmetry by a small amount, thus giving rise to the physical but small pion masses. In particular
$m_{\pi^+}^2$ receives an electromagnetic correction of order $\frac{\alpha_{EM}}{4\pi}\Lambda_{QCD}^2
\ll \Lambda_{QCD}^2$. We can try and think of an extension of the SM where the Higgs is a composite
Goldstone boson associated to some new strong dynamics at a scale $\Lambda_{\Strong}$. Among several others, the top Yukawa interaction (as it does not
involve derivatives of the Higgs field) breaks  the Goldstone symmetry explicitly.
 Then, replacing $\alpha_{EM}\to \alpha_{t}$ and $\Lambda_{QCD}\to \Lambda_{\Strong}$,
 we generically expect, in analogy with QCD,  $m_H^2 \sim \frac{\alpha_t}{4\pi} \Lambda_{\Strong}^2 $.
Since in this case $\Lambda_{\NP}\sim \Lambda_{\Strong} $ this is just eq.~(\ref{quadratic}),
and we are back to the LEP paradox. The Little Higgs \cite{littleH} is precisely a clever construction to avoid the
appearance of the lowest order contribution to $m_H^2$.  Consider indeed the expression for the mass of a 
Higgs pseudo-Goldstone boson,  to all order in the coupling constants
\be
m_H^2\,=\,\left (\,c_i\frac{\alpha_i}{4\pi}\,+\,c_{ij}\frac{\alpha_i\alpha_j}{(4\pi)^2}\,+\dots\right )\Lambda_{\Strong}^2\, .
\ee
We can think of these couplings $\alpha_i$ as external sources that transform non-trivially under the
Goldstone symmetry, thus breaking it, very much like an external electric field breaks the rotational invariance
of atomic levels. As in atomic physics,  the coefficients $c_i,\, c_{ij},\, \dots$ are controlled by the symmetry selection rules.
We can then in principle think of a clever choice of symmetry group and  couplings (thought of as external sources)
such that the Goldstone symmetry is partially restored when any single coupling $\alpha_i$ vanishes. 
In that situation only the combined effect of at least two distinct couplings $\alpha_i$ and $\alpha_j$ can 
destroy the Goldstone nature of the Higgs thus contributing a mass to it. 
The symmetry is said to be collectively broken,  $c_i=0$ and 
\be
m_H^2\sim (\frac{\alpha}{4\pi})^2 \Lambda_{\Strong}^2\, .
\ee
By this equation we then expect $\Lambda_{\Strong}\sim 10\, \tev$, which seems to be what we need to avoid the LEP paradox. 

 The general symmetry structure of LH models involves a global group $G_{\rm glo }$ broken down to a subgroup $ H_{\rm glo}$ with the Higgs doublet belonging to the Goldstone
 space ${G_{\rm glo}}/H_{\rm glo}$. Only a subgroup $G_{\rm loc}\subset { G_{\rm glo}}$ is gauged:
  gauge and  Yukawa interactions collectively realize the
  explcit breaking ${ G_{\rm glo}}\to G_{\rm loc} $. Therefore as a combination of spontaneous and explicit breaking only a gauge group
  $H_{\rm loc}\subset H_{\rm glo}$ survives between the fundamental scale $\Lambda_{\Strong}$ and the weak scale. Normally $H_{\rm loc}$ is just the electroweak group $G_{\rm weak}=SU(2)_L\times U(1)_Y$.
    In order to realize this structure, the field content of the SM must be clearly extended,
 and the many different ways of achieving that define a variety of Little Higgs models.
 One  feature of all these models is the presence of same spin partners for basically each SM field. When computing corrections to the Higgs mass, 
 these partners  enforce the selection rule $c_i=0$ by 
 cancelling the 1-loop quadratic divergent contribution of the corresponding SM field.
 For instance, in all models the left-handed top doublet $(t,b)_L$ is extended to at least a triplet 
$\chi_L=(t,b,T')_L$,
with $T'_L$ an up-type $SU(2)_L$ singlet; in the right-handed sector, along with $t_R$ and $b_R$, there is
then a new up-type quark  $ T'_R$. The field $\chi_L$ transforms as a triplet of some $SU(3)\subset { G_{\rm glo}}$. The ordinary Higgs boson arises as a (pseudo)-Goldstone
from the spontaneous breaking of $SU(3)$ down   to ordinary $SU(2)_L$. 
The triplet structure for third family fermions is a feature of the simplest models,
event though $G_{\rm glo}$ is strictly bigger than $SU(3)$.
For instance one  simple model is the so-called Littlest Higgs for which $ G_{\rm glo}=SU(5)$
and $ H_{\rm glo}=SO(5)$.

The gauge group can either be extended by adding extra group factors to
$G_{\rm{weak}}$ (product group models) or by embedding 
$G_{\rm {weak}}$ in  larger simple group (simple group models). For instance, within the latter class
the Simplest Little Higgs model  \cite{Schmaltz:2004de} has a weak gauge group
$SU(3)\times U(1)$.  The simplest product
group models instead, such as  the $SU(5)/SO(5)$ Littlest Higgs, have gauge group $SU(2)_1\times SU(2)_2\times U(1)_Y$.  
The role of the extra charged $W_H^\pm$ and neutral $Z_H$ is to cancel the 1-loop correction $\delta m_H^2 \sim \frac{\alpha_W}{4\pi}\Lambda_{\Strong}^2$
from SM vector bosons.
  
The partners of the SM states that are needed to enforce the LH mechanism naturally have  a mass of order
\be
m_{\rm{partners}}^2\sim \frac{\alpha}{4\pi} \Lambda_{\rm{\Strong}}^2= g^2 f^2\, ,
\ee
where I indicated by $\alpha=g^2/4\pi$ a generic coupling constant
and I used the qualitative relation  $\Lambda_{\Strong}\sim 4\pi f$
between the strong scale and the Goldstone decay constant $f$ (this is in  analogy with the relation between strong scale and $f_\pi$ in QCD). For $\Lambda_{Strong}\sim 10 \,\tev$, the partners then have  a mass in the TeV range. Notice that the presence of these new states with intermediate mass is necessary
for the LH mechanism to work.
\begin{figure}
\includegraphics[width=.8\textwidth]{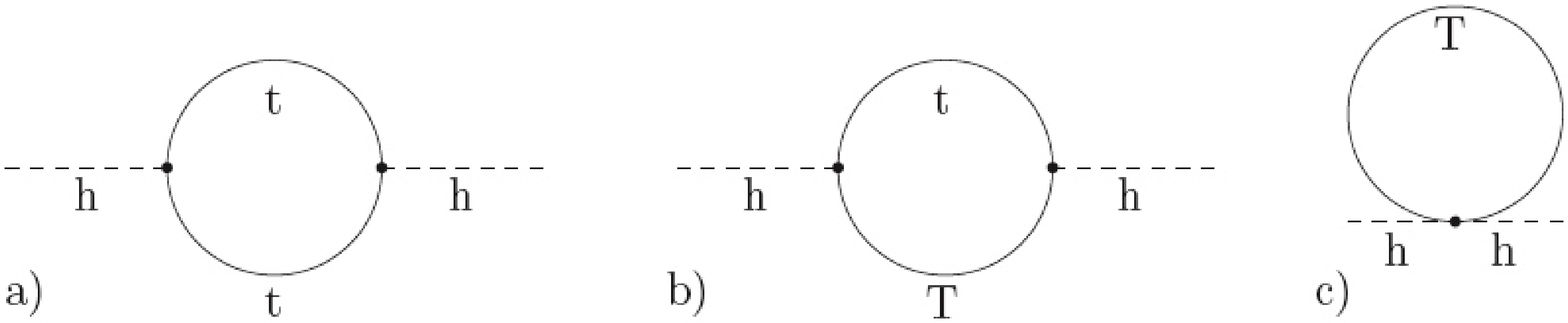}
\label{loops}
\end{figure}

 As already said, from the viewpoint of the low energy effective theory, the partner loops cancel the leading quadratic 1-loop correction to $m_H^2$. For instance in the top-quark sector
 the 3 diagrams in the figure add up  to a quadratic correction 
\be
\delta m_H^2=- \frac{3\Lambda_{\Strong}^2}{8\pi^2}\left (
\lambda_t^2+\lambda_T^2-2\frac{\lambda_T m_T}{f}\right )=0
\ee
thus implying a sum rule involving the top Yukawa, the $T$--$t$ mixing Yukawa, the heavy top partner mass $m_T$, and the Goldstone decay constant. An experimental validation of this sum rule would be a spectacular confirmation of the LH mechanism. The
cancellations among different diagrams are analogic to what happens in supersymmetry. The analogy goes indeed further, in that
logarithmic divergences do not cancel, and play a role in triggering
electroweak symmetry breaking. For instance in the Littlest Higgs model the $t$--$T$ sector gives rise to a negative correction
\be
\delta m_H^2 = -\frac{3}{8\pi^2}\lambda_t^2 m_T^2\ln \left (\frac{\Lambda}{m_T}\right )\, ,
\ee
completely analogous to the $t$--$\tilde t$ one in the MSSM.  

Those we just described  are undoubtedly attractive qualitative features for a theory of electroweak symmetry breaking. In the end, however, it is the comparison with the electroweak data that matters. In the LH models
there are two classes of contributions to effective operators.

The first class is associated to the yet unknown physics at the cut-off $\Lambda_{\Strong}$,
at which the Higgs is composite. It
necessarily gives rise to operators involving just the Higgs boson, where vector bosons appear only
through covariant derivatives. For $\Lambda_{\Strong}\sim 10\,\tev$ these effects are not in 
contradiction with the data. The situation would however be bad if light 
fermions
as well were composite at $\Lambda_{\Strong}$. This is because strong coupling would then demand
$c_i\sim 16\pi^2$ with $\Lambda_{\NP}=\Lambda_{\Strong}$ for 4-fermion contact interactions in eq.~(\ref{effective}). But with this normalization LEP2 data  imply $\Lambda_{\Strong}\gsim 50 \,\tev$ \cite{Barbieri:1999tm}.  Fortunately, fermion compositeness is not  a necessary requirement of LH models, although Higgs compositeness
requires some extra interactions in oder to give rise to the SM  Yukawa couplings. 

The second class of effects is mainly associated to the intermediate mass $\sim g f\sim 1 \, \tev$  vector bosons,
$W_H^\pm,\, Z^H,\,\dots$ In product group models, all such effects arise from the mixing between
heavy and light bosons. These models are therefore universal and all the new effects are faithfully parametrized 
by $\widehat S, \widehat T, Y, W$ \cite{Barbieri:2004qk,Marandella:2005wd}. Simple group models are not
universal because of the new current--current interactions associated to the extended gauge structure, but the bounds are roughly the same \cite{Marandella:2005wd,Han:2005dz}. 
 From the first analyses of electroweak data in LH models \cite{Csaki:2002qg} 
to the most recent and comprehensive ones \cite{Marandella:2005wd,Han:2005dz}, much work has been done.
In what follows I will briefly
discuss the results for product group models  as studied in ref.\cite{Marandella:2005wd}. One robust feature of these models is the contribution to $\widehat S$ and $W$;
in terms of the mass $m_{W_H}$ and gauge coupling $\alpha_H$ for the new vectors, this is just
\be
\label{SWLH}
\widehat S\,=\, \frac{m_W^2}{2m_{W_H}^2}\frac{1}{\sqrt{1-\frac{\alpha_W}{\alpha_H}}}\qquad\qquad
W\,=\, \frac{m_W^2}{2m_{W_H}^2}\frac{\frac{\alpha_W}{\alpha_H}}{\sqrt{1-\frac{\alpha_W}{\alpha_H}}}\, ,
\ee
while the contributions to $\widehat T$ and $Y$ are more model-dependent. However, especially thanks to LEP2, it is possible to strongly bound the model even by treating $\widehat T$ and $Y$ as free parameters.
 Notice that, by eq.~(\ref{SWLH}), it is the intermediate scale $m_{W_H}\sim g f \sim 1\,\tev$, instead of $\Lambda_{\Strong}=4\pi f$,
  that plays the role of the new physics scale $\Lambda_{NP}$: we are back to the LEP
paradox! In fact one may even say that the LH provides an explicit incarnation of the LEP paradox itself. 
By eq.~(\ref{SWLH}) the bound on $m_{W_H}$ and on $f$
 become  weaker as $\alpha_H$ gets larger. For $\alpha_H>0.3$
one gets  (with $95\%$ CL) $m_{W_H}>1.2\, \tev$, by keeping $\widehat T$ free,  and $m_{W_H}>1.6\, \tev$ for
$\widehat T = 0$. The direct bound on $m_{W_H},\alpha_H$ indirectly limits the mass of the top partner 
(via the bound on the LH decay constant $f$) roughly as
\be
m_T\,>\,\frac{1}{\sqrt {\alpha_H}}\,\tev\,.
\ee
We now see the LEP paradox in action. The
Higgs mass is dominated by quantum correction $\delta m_H^2\propto m_T^2$, and for a 
`normal'-size coupling $\alpha_H\lsim 0.1$ we must tune the Higgs mass to at least $5\%$ accuracy. Alternatively, tuning
is minimized, if we are willing to accept a  coupling $\alpha_H\sim O(1)$ on the verge of becoming strong.
While it does not seem technically unacceptable  to have such a large coupling at low energy,
 it may perhaps make things harder when trying to come up with a weakly coupled UV completion of the LH.
The need for slightly extreme choices of parameters is not limited  to product group models,
but also holds for simple group ones, although the discussion is somewhat different \cite{Marandella:2005wd,Han:2005dz}.
  Notice also that, in addition to the general tension with electroweak data,  
  specific models can have extra 
tuning \cite{Casas:2005ev}, for instance in association with the Higgs quartic coupling. 
I do not know whether it is fair to emphasize these more specific tunings. However I think it is fair to say  that for normally weak gauge couplings $\alpha_H\lsim 0.1$ the LH is not less tuned  than supersymmetry. 

 The basic problem involves the mixing between light and heavy vector bosons. However the cancellation of the leading quadratic correction to the Higgs mass does not rely on this mixing. In fact LH models have been constructed \cite{Cheng:2004yc} involving an extra discrete symmetry, T-parity,
with respect to which SM particles are even, while the heavy vector bosons are odd. This naturally forbids the mixing,
implying $\widehat S= W = 0$.  This would be a  great result, if it wasn't that with T-parity there necessarily appear
new and potentially disastrous loop corrections to 4-fermion contact terms. This is precisely what T-parity was asked to avoid!  These new loop effects are tamed provided a partner for each SM fermion, including
the light ones, is added with a mass around $500\,\gev$. This way, models with T-parity can probably  be made  technically less fine-tuned than models without it.  T-parity is a smart idea, but it is not clear to me if the extra complications it entails are worth the effort.

In the end, even if these models are somewhat cornered by LEP data,  it is only with the LHC that we will directly
test them. The top partners are likely to be the lightest and most accessible states, in view of tuning considerations.
$T$ is directly produced in $qb$ collisions   via the flavour mixing vertex $W^+ \bar T b$. The parameters $\lambda_T$ and $m_T$ are thus extracted from the measured rate and from the reconstructed mass. The remaining parameter
$f$, as well as $\alpha_H$, can be extracted from the Drell-Yan (DY) production and decay of the heavy vectors. 
 Notice that the large
$\alpha_H$ values that are favoured by low energy data suppress the coupling of $W_H$ to light particles, thus leading
to a suppression in the DY cross section. One can still conclude that, in case of a discovery up to $m_T<2.5 \, \tev$ and $m_{W_H}<3\, \tev$, the sum rule eq.~(\ref{quadratic}) can likely be tested within 10\% accuracy \cite{Perelstein:2003wd}.

\subsection{$H\sim A_5$ or Higgs as  `holographic'  Goldstone boson} 
This is also a pretty `old' idea \cite{manton,hosotani} on which, again,  progress was made in recent years (see for instance ref.~\cite{many}) thanks
to the use of new concepts such as  branes, warping, deconstruction, etc.
The basic remark is that when the gauge group $G_{\extra}$ breaks down to
$G_{\weak}$ by some clever compactification, the extra-dimensional polarizations $A^\alpha_5,\,A^\alpha_6,\,\dots$, associated to the generators $T_\alpha\in G_{\extra}/G_{\weak}$, are massless at tree level. Very much as for the LH models, one can build models
where $G_{\extra}/G_{\weak}$ contains the SM Higgs doublet. The extra dimensional gauge symmetry then forbids the presence of local contributions to the mass of such a Higgs boson,
implying that all the contributions to $m_H^2$ must be associated to non-local, ie. finite,
quantum corrections. 

\begin{figure}
\includegraphics[width=.8\textwidth]{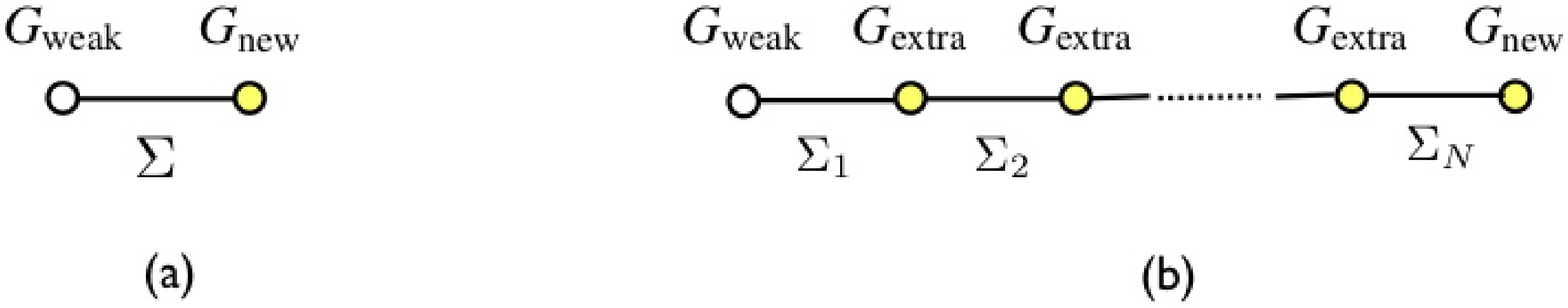}
\label{moose}
\end{figure}
These models are indeed closely related to a large class of LH. These are the so-called moose models, which can be represented by diagrams where the
dots indicate gauge group factors, while the links indicate scalar fields with quantum numbers
under the two gauge dots they connect. A simple LH moose, depicted in figure (a), 
involves one extra gauge group factor $G_{new}$ admitting $SU(2)\times U(1)$ as a subgroup.
The link field  $\Sigma$ represents  
the set of Goldstone bosons associated to the breaking of the global symmetry group $G_{\rm glo}=G_{\extra}\times G_{\extra}$ down to just $G_{\extra}$.  The gauge factors 
 $G_{\weak}$ and $G_{\rm new}$ are embedded into each distinct $G_{\extra}$ factor of $G_{\rm glo}$.
The Goldstone field $\Sigma$ breaks $G_{\weak}\times G_{\rm new}\to G_{\weak}$, and the uneaten
Goldstones $\in G_{\extra}/G_{\rm new}$ make up the Higgs doublet. Notice that this construction
realizes collective symmetry breaking: in the limit of vanishing  gauge coupling
for any individual dot (either $G_{\weak}$ or $G_{\rm new}$), $\Sigma$ 
becomes an exact Goldstone. Now, one may imagine
repeating this structure by adding $N$ intermediate dots with gauge group $G_i\equiv G_{\extra}$, linked by replicas of $\Sigma$, as shown in the figure. In the limit $N\gg 1$,
this linear structure truly approximates a 5-dimensional theory, with gauge group
$G_{extra}$, broken down to respectively $G_{\weak}$ and $G_{\rm new}$ at 
each boundary. The scalars $\Sigma_i$ play the role of $A_5(i)$, which makes the connection
between $H\sim A_5$ and LH fully manifest. The moose diagram is called a 
{\it deconstruction} of  the 5th dimension \cite{Arkani-Hamed:2001ca}. The Higgs mass is calculable at 1-loop, as in any LH model, but now the new states that cut-off the quadratic divergence are nicely
interpreted as the Kaluza--Klein replicas of the SM fields. So we roughly have
\be
\label{mhKK}
m_H^2\sim \frac{3\lambda_t^2}{16 \pi^2} m_{KK}^2+\dots
\ee

At the classical level one may think of achieving the continuum limit by sending $N\to\infty$.
At the quantum level, however, that does not make sense, since 5-dimensional gauge theories have a UV cut-off
$\Lambda_{\Strong}$ which sets a minimal length scale $1/\Lambda_{\Strong}$.
Indeed,  studying the spectrum of the deconstructed theory one gets
\be
\label{nkk}
N\sim \Lambda_{\Strong}/m_{KK}\sim\Lambda_{\Strong} R
\ee
so that  $N$ truly has a physical interpretation as the number of
weakly coupled  Kaluza--Klein resonances below the cut-off $\Lambda_{\Strong}$ (the KK levels  
  are more or less equally spaced by an amount $m_{KK}$). 
  The simple construction we have just sketched  displays, perhaps roughly,
  some aspects of a more general  idea, {\it holography}, born within string theory \cite{maldacena}, but more   and more influential in model building and phenomenology (see e.g. ref. \cite{Gherghetta:2006ha}): weakly coupled 5D theories
  can be alternatively viewed as purely 4D  theories with a large number of states $N$.  Moreover, the more weakly coupled the 5D description (the larger $\Lambda_{\Strong}$)  the larger $N$.

  Among the various realizations of $H\sim A_5$,  the arguably most interesting one \cite{Agashe:2004rs} was  obtained within the  Randall--Sundrum (RS) model \cite{Randall:1999ee}. In the RS model the 5th dimensional interval  $y=[0,R]$ is warped and the metric is
  $ds^2=e^{-2y/L}dx_\mu dx^\mu +dy^2$. The length $L$ characterizes the distance 
  along $y$ beyond which curvature effects are important.   The warp factor $e^{-y/L}$
  describes the energy red-shift of any process taking place at $y$, relative
  to the same process taking place at $y=0$. This is conceptually analogous to the 
  relative red-shift of light emitted in a given atomic transition by atoms sitting
  at different heights in the gravitational field of the Earth. However, unlike on Earth, in the RS metric the curvature of space-time is large. The red-shift is then huge, and can be used to explain the Big Hierarchy problem. Indeed in the RS model the effective 4-dimensional force is 
  mediated by a massless graviton localized near $y=0$, and therefore the effective $M_{\Planck}$   is not red-shifted. However the lightest Kaluza-Klein states, for all fields,  are localized  near $y=R$ and their mass is red-shifted by a factor $e^{-R/L}$:
 \be
 \label{bighie}
 \frac{m_{KK}}{M_{\Planck}}\sim e^{-R/L}\,.
 \ee
 If one succeeds in associating $m_{KK}$ to the weak scale, then 
the exponential explains the Big Hierarchy for a fairly small  radius $R/L\sim 35$.
In the model of ref.~\cite{Agashe:2004rs} the Higgs is basically the zero model of some
components of $A_5$: $H\,=\,\int_0^R\, A_5 \,dy$. Its mass, generated at 1 loop,
is of the form in eq.~(\ref{mhKK}) as expected in any $H\sim A_5$ construction.
The peculiarity of this model is then that the calculability of $m_H$ is combined with
a solution of the Big Hierarchy problem. Unlike most LH theory, thanks to the embedding in the RS geometry, the model in ref.~\cite{Agashe:2004rs} gives
a valid description of physics up to energies of the order of $M_{\Planck}$. In this sense it
can be considered a serious competitor of supersymmetry. As in supersymmetry,
the extrapolation to the Planck scale is rather constraining:
\begin{itemize}

\item There are KK resonances for each SM particle.

\item Perturbativity of the three SM gauge couplings up to the Planck scale implies $N\lsim 10$, 
see eq.~(\ref{nkk}). By converse this bound implies that the coupling among KK modes is pretty strong: $g_{KK}\sim \frac{4\pi}{\sqrt N}$. The KK states behave like the resonances of a strongly coupled 4-dimensional field theory.

\item The quark and lepton mass spectrum can be nicely explained via their localization in 5D,
while implementing a GIM mechanism to suppress FCNC's \cite{Grossman:1999ra}.

\item The right-handed top $t_R$, unlike the other SM states, strongly interacts with the KK modes. From the 4-dimensional perspective the interpretation is that $t_R$ is composite.

\item The electroweak constraints are similar to the LH as far as oblique corrections are concerned:
they require  about $10\%$ tuning corresponding to  a bound  $m_{W_H}> 2.5\,\tev$ on the mass of the lightest vector KK mode, slightly stronger than for LH.
However significantly stronger bounds are here associated to corrections to the $Z b\bar b$ vertex  \cite{Agashe:2005dk}. They lead to a bound of about $4\, \tev$ on the mass of the top KK partners, thus implying a need for fine tuning at the few per cent level. These stronger bounds,
unlike the more robust ones from $\widehat S$, may however be a peculiarity of the specific model, and some possibilities to overcome them are outlined in ref. \cite{Agashe:2005dk}.
\end{itemize}

One last item concerns gauge unification, which in some leading, naive, approximation
 works very well, and in a novel way, totally alternative to what was thought so far \cite{Agashe:2005vg}. The beta
function are indeed not just modified by the addition of the contribution of new states, but also by the subtraction of the contribution of the Higgs $H$ and the right-handed top $t_R$, which are by all means composite states just above the weak scale. The problem, however, is that higher order uncalculable effects are very important unless $N\sim O(1)$, which would drastically limit
the overall calculability \cite{Agashe:2005vg}. So, while the idea of unification by subtraction is new and interesting, it does not have yet a realization that can computationally compare to the fully weakly
coupled supersymmetric unification.

5D models or moose models, can also be used to construct 
partially calculable Higgsless theories \cite{Csaki:2003dt}. This corresponds to choosing $G_{\rm new}$ and its embedding in $G_{\extra}$ so that the combined effect of the two boundaries is to break 
$G_{\extra}$ directly to electric charge $U(1)_Q$ (in the 1-link moose limit, diagram (a),
the link field breaks $G_{\weak}\times G_{\rm new}$ to $U(1)_Q$.).
 These models are very ambitious since,  unlike in models with a Higgs field, 
 the ratio $(m_Z/m_{W_H})^2$ is fixed in any given construction to be of order $g^2 N/16\pi^2$,
and it is not tunable. This makes it harder to pass the electroweak precision tests:
either small $N$ is chosen \cite{Barbieri:2004qk}, implying unacceptably strong coupling, or the simplicity of the idea
must be spoiled by extra complications \cite{Cacciapaglia:2005pa}. Moreover it is not yet clear
if non--universal effects such as  $Z\to b\bar b$ can be fully kept under control.

\section{Anthropic approach to hierarchy problem(s)}
The ideology underlying model building attempts, such as  the ones I described so far,
is that the measured parameters of the SM must be pointing toward a unique fundamental description of Nature. If that description is not perverse, any apparent tuning within the SM should not look
so within the more fundamental description. Thus we must look for theories that 
effortlessly explain the value of the weak scale, $\theta$-QCD, etc. The anthropic approach to physics, 
and to the hierarchy problems in particular, follows a different ideology, which could 
be based on a multiverse assumption:
\begin{itemize}
\item Our local universe represents  but  a small
region of a multiverse in which some, perhaps all, physical parameters vary from region to region.
\end{itemize}
According to the multiverse assumption, the value of some physical quantities, which we so far
considered a fundamental property of Nature, may instead have a purely environmental origin.
One standard example of an environmental quantity is the radius
of Earth's orbit around the Sun: while not fundamental, its value is pretty constrained
by the prior that  the Earth has  a hospitable atmosphere with the presence of liquid water.
The anthropic principle was for long considered powerless by the great majority,
until Weinberg in 1987 \cite{Weinberg:1987dv}  applied it to the cosmological 
 constant $\Lambda_{\cosm}$ , thus providing a radically different viewpoint on the least understood 
 of all hierarchy problems. Weinberg's assumed that (Structure Principle) 
 \begin{enumerate}
\item $\Lambda_{\cosm}$ is not a fundamental quantity.
\item The only environmental constraint on $\Lambda_{\cosm}$ is that it be small enough to allow the formation of galaxies.
\end{enumerate}
Weinberg then argued that, if the distribution of values of $\Lambda_{\cosm}$ is reasonably smooth, then the most natural expectation is that $\Lambda_{\cosm}$ be of the same order of magnitude as, or not much smaller than, the critical value $\Lambda_{c}$ below which galaxies can form. Then, 
when there was still no observational indication that $\Lambda_{\cosm}\not = 0$,
 Weinberg  predicted a likely value $\Lambda_{\cosm }\sim \Lambda_c \sim 100 \rho_c$,
 where $\rho_c$ is the critical density for the closure of the Universe. 
  The computation was later refined into roughly $\Lambda_c\sim 10 \rho_c$ \cite{Martel:1997vi}.
In the meanwhile the Type IA Supernovae data \cite{Riess:1998cb} had established
the presence of a negative pressure energy density component, compatible with a cosmological constant $\Lambda_{\cosm}\simeq 0.7 \rho_c$.  It is quite remarkable that Weinberg's logic 
correctly predicts, to within an order of magnitude,  a mysterious quantity like $\Lambda_{\cosm}$,
which is otherwise apparently tuned by 120 orders of magnitude ($ \Lambda_{\cosm}/M_{\Planck}^4\sim 10^{-120}$).

Further to the success of the Structure Principle, the anthropic viewpoint has recently also been reinforced by advances is string theory,  indicating the existence of a tremendous multitude of different vacua, forming what is called the Landscape. The universe would then be a multiverse
with each different region (subuniverse) sitting at a different vacuum out in the Landscape.
That and the frustration with standard approaches have stimulated
the use of the anthropic viewpoint on the electroweak hierarchy problem. Ref.~\cite{Agrawal:1997gf}
introduced what is now called the Atomic Principle, according to which the Fermi scale is an environmental quantity whose value is nailed by the request that complex chemistry (atoms)
exists. Remarkably the Atomic Principle sets an upper bound on $\langle H\rangle$, which is only about 5 times its experimental value. The Atomic Principle was later applied to the MSSM \cite{Arkani-Hamed:2004fb,Giudice:2004tc}
under the assumption that the soft terms, and thus the weak scale, are environmental quantities,
and with the additional request that the lightest supersymmetruc particle (LSP), a neutralino, provide the Dark Matter  of the Universe. The resulting scenario, dubbed Split Supersymmetry, features superheavy (even up to  $10^{13}\,\gev$)
squarks, leptons and one combination of Higgs scalars, while the charginos and the neutralinos 
have a mass which `accidentally' ends up close to the weak scale in order to have the right
amount of relic LSP. Remarkably, although the superspectrum is split, the successful
unification of gauge couplings is mantained, as in supersymmetry that is mostly due to the 
contribution of Higgsinos and gauginos. Moreover the set up is rather predictive. In particular
the gluino decays very slowly, via the virtual exchange of the heavy squarks, giving rise,
 over a significant portion of parameter space,  to distinctive displaced
 vertex events. Split Supersymmetry has been the subject of a great amount of work in the last year. It  would be fair and worthwhile to review this work, but unfortunately I do not have enough time.
 In the remaining part of my talk I would instead like to present a new, anthropic, viewpoint on
 the fine-tuning problem of the MSSM \cite{agr}.
 
 \subsection{Back to Supersymmetry}
Let us go back to the well known cartoon of EWSB by RG evolution in
supersymmetry, shown in the figure. The Higgs mass parameter $m_H^2$ starts positive up at the 
Planck scale and is driven negative below some RG scale $Q_{\crit}$ by RG contributions,
mostly due to the stops, until the running is frozen at the typical scale 
of sparticle masses $m_{SUSY}$. The physical value of the Higgs mass is then approximatively

\begin{figure}
\includegraphics[width=.7\textwidth]{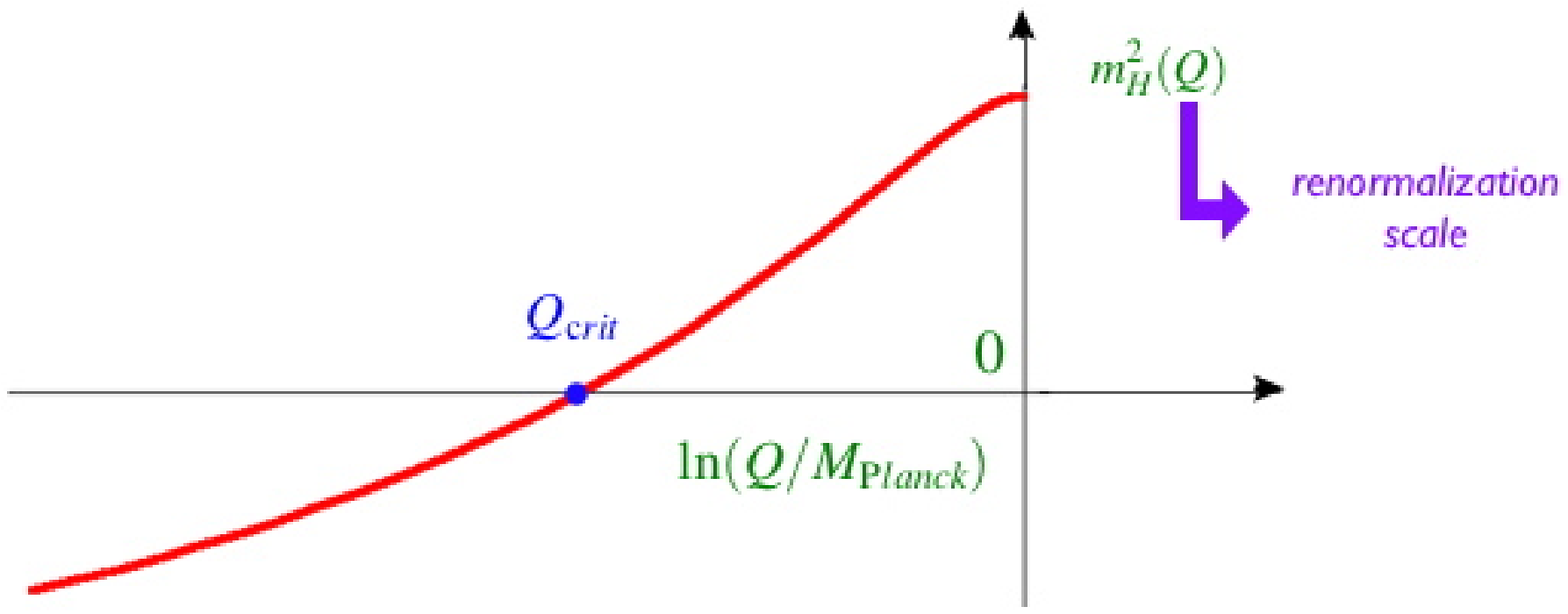}
\label{crit}
\end{figure}

\be
m_H^2\vert_{\phys}\,=\,m_H^2(Q=m_{\SUSY})
\ee
 Now, $Q_{\crit}$ 
is associated to a dimensional transmutation, and is expected to differ
significantly from both  $m_{\SUSY}$ and $M_{\Planck}$.
A generic expectation is $m_{\SUSY}\ll Q_{\crit}\ll M_{\Planck}$ so that by the time the running is frozen at $Q\sim m_{\SUSY}$ the Higgs mass is large and negative (cf. eq.~(\ref{zstop}))
\be
 m_H^2\vert_{\phys}\,\sim\,
-m_{\tilde t}^2\,\sim\,-m_{\SUSY}^2\,.
\ee
As we explained already, this is unfortunately not the situation favoured by the data. The data 
favour  instead $-m_H^2\vert_{\phys}\,\ll\, m_{\tilde t}^2$, which by direct glance at the
figure is equivalent to
\be
\label{why}
m_{\SUSY} \simeq Q_{\crit}.
\ee
An alternative way of phrasing the fine-tuning problem of supersymmetry  then is  to ask:
Why should two totally unrelated parameters like $m_{\SUSY}$ and $Q_{\crit}$ 
almost coincide?  Given the present constraints, if Supersymmetry is discovered at the LHC,
we will almost certainly have to ask ourselves this question. Let me try and give an answer
right now.

Let us assume that the overall SUSY mass scale $m_{\SUSY}$ is environmental.
More precisely, let us assume that up at the Planck scale the various soft parameters are given by
\be
m_i= c_i\, m_{\SUSY}\, ,
\ee
with the $c_i$ fixed everywhere throughout the multiverse, while $m_{\SUSY}$ varies. Let
us also assume that all the other dimensionless gauge and Yukawa couplings are fixed at the
Planck scale. Notice that under these conditions $Q_{\crit}$ is also fixed, as it depends only on $M_{\Planck}$, $c_i$, and the other dimensionless couplings, but not on  $m_{\SUSY}$.
Two possibilities for the patch of Universe we live in are then given
\begin{enumerate}
\item $m_{\SUSY}> Q_{\crit}$, in which case $m_H^2\vert_{\phys}>0$, implying $\langle H\rangle = 0$.
\item $m_{\SUSY}<Q_{\crit}$, in which case $m_H^2\vert_{\phys}<0$, implying $\langle H\rangle \not = 0$.
\end{enumerate}
It is pretty clear we do not live in region 1, and in fact it is not  even sure if in region 1
there can exist anyone to ask this question \cite{Agrawal:1997gf,Arkani-Hamed:2005yv}.
Now, compatibly with the prior that we must live in region 2, what is the most likely value
we expect $m_{\SUSY}$ to have? The problem is phrased in complete analogy
with Weinberg's approach to the cosmological constant, with $\langle H\rangle \not = 0$
replacing the datum that galaxies exist. Then, under the assumption that the distribution
of $m_{\SUSY}$ is reasonably flat and featureless, and, which is quite likely, not peaked at $m_{\SUSY}=0$, we expect $m_{\SUSY}\sim Q_{\crit}$. A small Higgs mass parameter is then
obtained through the brief running from $Q_{\crit}$ to $m_{\SUSY}$
\be
\frac{ m_H^2\vert_{\phys}}{ m_{\tilde t}^2}\sim \frac{3\lambda_t^2}{2\pi^2}\,\ln(m_{\SUSY}/Q_{\crit})\sim -\frac{3\lambda_t^2}{2\pi^2}\ll 1\, .
\ee
To be more precise let me assume   the number $N(m)$ of vacua with $m_{\SUSY}< m$ grows like $m^n$. The 
prior $m_{SUSY}< Q_{crit}$ leads to a conditional probability giving the average
$\langle \ln(m_{\SUSY}/Q_{\crit})\rangle \sim -1/n$,
so that the expectation is
\be
\label{ratio}
\frac{ m_H^2\vert_{\phys}}{ m_{\tilde t}^2}\sim -\frac{3\lambda_t^2}{2\pi^2}\times \frac{1}{n}\, .
\ee
Notice that the loop factor $3\lambda_t^2/2\pi^2\sim 0.15$, while it helps to explain the
little hierarchy problem in supersymmetry, falls short to explain it completely. Indeed one
can imagine field theoretic landscapes \cite{agr} where $n$ is somewhat bigger than 1, say $O({\rm a \,few})$ but not much
bigger (For instance if there are $O(10^{500})$ vacua, as perhaps suggested by string theory, and if $m_{\SUSY}$ can range up to $M_{\Planck}$, then $n\lsim 30$). So it is reasonable
for the ratio in eq.~(\ref{ratio}) to be between $0.01$ and $0.1$ but not much smaller,
thus providing an argument  why supersymmetry should be elusive at LEP but not at the LHC.
Of course there has been a price to pay. Supersymmetry  looks tuned because throughout the Lanscape it is  much more likely  to be in the region with  $\langle H \rangle  = 0$ 
 than in the region  $\langle H \rangle\not = 0$ : the most likely points with
 $\langle H \rangle \not = 0$ are then  close to the boundary of the two regions, where a little hierarchy is present.
 
 Now, what does one do with an argument like the above?
 Can it be falsified? It certainly can. It predicts that  $m_H^2$ will cross zero immediately
 above the supersymmetric threshold as we run the soft parameters up in energy. Now,
 although less typical, or even more tuned, there are choices of parameters where this does not happen, for instance when the beta function for $m_H$ has a zero at the weak scale. Such 
 values of the soft masses would rule out this scenario (although it would probably be hard,
 given the precision with which masses are measured at the LHC, to quickly reach a conclusion).
 Another situation that would rule out this scenario is that in which $m_H^2$ becomes negative
 at some high RG scale, as  it would happen in small deformations of gaugino mediation.
In the end, is the possibility to falsify this scenario so exciting? Probably not, as, if supersymmetry is discovered, it will very likely look like that. I think the main lesson is that fine-tuning in
supersymmetry, once we discover it, could be telling us something fundamental 
about the statistics of vacua and the nature of soft terms up at the Planck scale.
 
 \section{Summary}
In recent years there have been many new proposals of calculable electroweak symmetry breaking,
all trying to account for the baffling absence of new signals at LEP/SLC.
In practically all the examples there are two separate energy scales
\begin{itemize}
\item $\Lambda_{\NP}\,\sim\, 1\,{\tev}$, at which lay  particles that regulate the Higss mass divergence.
\item $\Lambda_{\Strong }\,\sim\, 10\,{\tev }$,  which describes the underlying new (strong) dynamics.
\end{itemize}
In all the models there exists already some tension with electroweak precision tests mostly as a consequence of the need for states at a relatively low scale $\Lambda_{\NP}\,\sim\, 1\,{\tev}$. 
In fact it is  fair to say that models such as the Little Higgs or the Holographic Goldstone boson
are explict incarnations of the LEP paradox.
The tension is not dramatic yet and can be relaxed at the price of some extra complications
(large gauge couplings or $T$-parity in Little Higgs models), so perhaps one should not worry too much. After all the LHC will directly test, in many of  these models,
a sizeable  portion of the parameter space, which is not constrained even indirectly by LEP. In particular the LHC will test the lower layer structure up to $\Lambda_{NP}\sim 3\, \tev$. The comparison
of these new approaches to SUSY is a fair exercise.
But one should be careful not to compare apples and oranges.
Supersymmetry provides a weakly coupled calculable description
for physics up to the Planck scale.  The extrapolation is rather constraining and  thus accounts for a good portion of the tuning that is needed in the MSSM. If we set ourselves the  less ambitious  goal  of finding a theory of electroweak symmetry breaking valid only to slightly above the weak scale as is done in most Little Higgs 
models, then supersymmetry would look less tuned. The 5D supersymmetric model presented in Ref.~\cite{Barbieri:2002sw} is an illustration of
that possibility. On the other hand the holographic Higgs Goldstone model \cite{Agashe:2004rs} can be  extrapolated up to the Planck scale, which
makes it fully comparable to the MSSM, and also very constrained \cite{Agashe:2005dk}! More concretely, perhaps, the new models compare reasonably well with supersymmetry as Dark Matter is concerned.
But this is largely due to the fact that any stable relic with weak scale
annihilation cross-section is a potentially good  Dark Matter candidate.
In the models at hand the stability of the relic follows from a discrete symmetry, for instance T-parity in LH or KK parity in the 5D models,
precisely has it follows from R-parity in supersymmetry.  On the other hand,
the neatness of gauge unification in supersymmetry is in my opinion not matched by any of the new models, although new intriguing twists have emerged \cite{Agashe:2005vg}.

The biggest novelty of the last year is however that the anthropic principle has finally made it to the gauge hierarchy problem. Weinberg's impressive anthropic explanation of the size of the cosmological constant, together with the lack of a fully natural theory
of electroweak symmetry breaking, is perhaps a serious indication that we do live in a multiverse
of vacua. How do we proceed if that is the case? We can certainly toy with the Landscape and try to come up with alternative solutions to the problems of particle physics. In this respect I illustrated a new viewpoint on the supersymmetric fine-tuning problem. With Split Supersymmetry the anthropic approach has even
materialized into a  cleverly predictive model. However I find it worrysome that with
the anthropic approach we are working with assumptions that are very hard, probably impossible, to test. The multiverse theory may become the greatest revolution after Copernicus, but will we ever test it?

Luckily a less speculative era will start in a couple of years, as the LHC will start to unravel  under our eyes
 the riddle of the weak scale.
 
 \vskip 1.0truecm
 
 I would like to thank Nima Arkani-Hamed, Kaustubh Agashe, Riccardo Barbieri, Roberto Contino, Gian Giudice, Thomas Gregoire, Christophe Grojean, Alex Pomarol, Martin Schmaltz, Claudio Scrucca, Alessandro Strumia and Raman Sundrum for many instructive discussions.


\begin{thebibliography}{99}
\bibitem{Barbieri:2000gf}
  R.~Barbieri and A.~Strumia,
  arXiv:hep-ph/0007265.
\bibitem{deJong:2005mk}
  S.~de Jong, these proceedings,
  arXiv:hep-ex/0512043.
\bibitem{Buchmuller:1985jz}
  W.~Buchmuller and D.~Wyler,
  Nucl.\ Phys.\ B {\bf 268}  621 (1986).
\bibitem{Barbieri:1999tm}
  R.~Barbieri and A.~Strumia,
  Phys.\ Lett.\ B {\bf 462}, 144 (1999).
  \bibitem{romanino} A. Romanino, private communication.
  \bibitem{agr} N. Arkani-Hamed, G.F. Giudice, R. Rattazzi, paper in preparation.
\bibitem{Bastero-Gil:2000bw}
  M.~Bastero-Gil, C.~Hugonie, S.~F.~King, D.~P.~Roy and S.~Vempati,
  Phys.\ Lett.\ B {\bf 489}, 359 (2000).
  \bibitem{tech}M.~E.~Peskin and T.~Takeuchi,
Phys.\ Rev.\ Lett.\ {65}  964 (1990);
M.~Golden and L.~Randall,
Nucl.\ Phys.\ {B361} 3 (1990);
B.~Holdom and J.~Terning,
Phys.\ Lett.\ {B247}  88 (1990).
\bibitem{Barbieri:2004qk}
  R.~Barbieri, A.~Pomarol, R.~Rattazzi and A.~Strumia,
  Nucl.\ Phys.\ B {\bf 703}, 127 (2004).
\bibitem{Grinstein:1991cd}
  B.~Grinstein and M.~B.~Wise,
  Phys.\ Lett.\ B {\bf 265}  326 (1991).
  \bibitem{pt}
M.~E.~Peskin and T.~Takeuchi,
Phys.\ Rev.\ D {46} 381 (1992) .
\bibitem{alba}
G.~Altarelli and R.~Barbieri,
Phys.\ Lett.\  {B253}  161 (1991).
\bibitem{Kaplan:1983fs}
  D.~B.~Kaplan and H.~Georgi,
  Phys.\ Lett.\ B {\bf 136}, 183 (1984).
  \bibitem{littleH}
N.~Arkani-Hamed, A.~Cohen and H.~Georgi,
 Phys.\ Lett.\ { B513}  232 (2001);
 N.~Arkani-Hamed, A.~Cohen, E.~Katz and A.~Nelson,
JHEP { 07} 034 (2002);
N.~Arkani-Hamed, A.~G.~Cohen, E.~Katz, A.~E.~Nelson, T.~Gregoire and J.~G.~Wacker,
JHEP {0208}  021 (2002); for a nice review see
  M.~Schmaltz and D.~Tucker-Smith,
  arXiv:hep-ph/0502182.
\bibitem{Schmaltz:2004de}
  M.~Schmaltz,
  JHEP {\bf 0408}, 056 (2004).
\bibitem{Csaki:2002qg}
  C.~Csaki, J.~Hubisz, G.~D.~Kribs, P.~Meade and J.~Terning,
  Phys.\ Rev.\ D {\bf 67}, 115002 (2003).
\bibitem{Marandella:2005wd}
  G.~Marandella, C.~Schappacher and A.~Strumia,
  Phys.\ Rev.\ D {\bf 72} 035014 (2005).
\bibitem{Han:2005dz}
  Z.~Han and W.~Skiba,
  Phys.\ Rev.\ D {\bf 72}  035005 (2005).
\bibitem{Casas:2005ev}
  J.~A.~Casas, J.~R.~Espinosa and I.~Hidalgo,
  JHEP {\bf 0503}, 038 (2005).
  
\bibitem{Cheng:2004yc}
  H.~C.~Cheng and I.~Low,
  JHEP {\bf 0408}, 061 (2004); I.~Low,
  JHEP {\bf 0410}, 067 (2004).
 
\bibitem{Perelstein:2003wd}
  M.~Perelstein, M.~E.~Peskin and A.~Pierce,
  Phys.\ Rev.\ D {\bf 69}, 075002 (2004);
  T.~Han, H.~E.~Logan and L.~T.~Wang,
  arXiv:hep-ph/0506313.
   \bibitem{manton} N.~S.~Manton, Nucl.\ Phys.\ B {\bf 158} (1979) 141; D.~B.Fairlie,
  Phys.\ Lett. \ B{\bf 82} 97 (1979).
  \bibitem{hosotani} Y.~Hosotani, Ann.\ Phys. \ {\bf 190}  285 (1989).
  \bibitem{many}  H.~Hatanaka, T.~Inami and C.~S.~Lim,
  Mod.\ Phys.\ Lett.\ A {\bf 13}, 2601 (1998); G.~R.~Dvali, S.~Randjbar-Daemi and R.~Tabbash,
  Phys.\ Rev.\ D {\bf 65}, 064021 (2002); 
 I.~Antoniadis, K.~Benakli and M.~Quiros,
  New J.\ Phys.\  {\bf 3}, 20 (2001);
  C.~Csaki, C.~Grojean and H.~Murayama,
  Phys.\ Rev.\ D {\bf 67}, 085012 (2003);
   C.~A.~Scrucca, M.~Serone and L.~Silvestrini,
  Nucl.\ Phys.\ B {\bf 669}, 128 (2003).
   
\bibitem{Arkani-Hamed:2001ca}
N.~Arkani-Hamed, A.~G.~Cohen and H.~Georgi,
Phys.\ Rev.\ Lett.\  {\bf 86}, 4757 (2001); H.~C.~Cheng, C.~T.~Hill, S.~Pokorski and J.~Wang,
Phys.\ Rev.\ D {\bf 64}, 065007 (2001).
\bibitem{maldacena} J.~Maldacena,  Adv. Theor. Math. Phys.  {\bf 2}
 231 (1998).
\bibitem{Gherghetta:2006ha}
  T.~Gherghetta,
  arXiv:hep-ph/0601213.
\bibitem{Agashe:2004rs}
  K.~Agashe, R.~Contino and A.~Pomarol,
  Nucl.\ Phys.\ B {\bf 719}, 165 (2005);
   R.~Contino, Y.~Nomura and A.~Pomarol,
  Nucl.\ Phys.\ B {\bf 671}, 148 (2003).
\bibitem{Randall:1999ee}
  L.~Randall and R.~Sundrum,
  Phys.\ Rev.\ Lett.\  {\bf 83}, 3370 (1999).
\bibitem{Grossman:1999ra}
  Y.~Grossman and M.~Neubert,
  Phys.\ Lett.\ B {\bf 474}, 361 (2000);  T.~Gherghetta and A.~Pomarol,
  Nucl.\ Phys.\ B {\bf 586}, 141 (2000); S.~J.~Huber and Q.~Shafi,
  Phys.\ Lett.\ B {\bf 498}, 256 (2001).
\bibitem{Agashe:2005dk}
  K.~Agashe and R.~Contino,
  arXiv:hep-ph/0510164.
\bibitem{Agashe:2005vg}
  K.~Agashe, R.~Contino and R.~Sundrum,
  Phys.\ Rev.\ Lett.\  {\bf 95}, 171804 (2005).
\bibitem{Csaki:2003dt}
  C.~Csaki, C.~Grojean, H.~Murayama, L.~Pilo and J.~Terning,
  Phys.\ Rev.\ D {\bf 69}, 055006 (2004).
\bibitem{Cacciapaglia:2005pa}
  G.~Cacciapaglia, C.~Csaki, C.~Grojean, M.~Reece and J.~Terning,
  Phys.\ Rev.\ D {\bf 72}, 095018 (2005)
\bibitem{Weinberg:1987dv}
  S.~Weinberg,
  Phys.\ Rev.\ Lett.\  {\bf 59}, 2607 (1987).
\bibitem{Martel:1997vi}
  H.~Martel, P.~R.~Shapiro and S.~Weinberg,
  Astrophys.\ J.\  {\bf 492}, 29 (1998).
\bibitem{Riess:1998cb}
  A.~G.~Riess {\it et al.}  [Supernova Search Team Collaboration],
  Astron.\ J.\  {\bf 116}, 1009 (1998);
   S.~Perlmutter {\it et al.}  [Supernova Cosmology Project Collaboration],
  Astrophys.\ J.\  {\bf 517}, 565 (1999).
\bibitem{Agrawal:1997gf}
  V.~Agrawal, S.~M.~Barr, J.~F.~Donoghue and D.~Seckel,
  Phys.\ Rev.\ D {\bf 57}, 5480 (1998).
\bibitem{Arkani-Hamed:2004fb}
  N.~Arkani-Hamed and S.~Dimopoulos,
  JHEP {\bf 0506}  073  (2005).
\bibitem{Giudice:2004tc}
  G.~F.~Giudice and A.~Romanino,
  Nucl.\ Phys.\ B {\bf 699}, 65 (2004)
  [Erratum-ibid.\ B {\bf 706}, 65 (2005)];  N.~Arkani-Hamed, S.~Dimopoulos, G.~F.~Giudice and A.~Romanino,
  Nucl.\ Phys.\ B {\bf 709}, 3 (2005).
\bibitem{Arkani-Hamed:2005yv}
  N.~Arkani-Hamed, S.~Dimopoulos and S.~Kachru,  arXiv:hep-th/0501082.
\bibitem{Barbieri:2002sw}
  R.~Barbieri, L.~J.~Hall, G.~Marandella, Y.~Nomura, T.~Okui, S.~J.~Oliver and M.~Papucci,
  Nucl.\ Phys.\ B {\bf 663}, 141 (2003).
\end{thebibliography}
\end{document}